%% file: 00_main_Jeevanandam_Jain_IEEE_ITSC_2025.tex
\documentclass[letterpaper, 10pt, conference]{IEEEtran}
\IEEEoverridecommandlockouts
\usepackage{cite}
\usepackage{amsmath,amssymb,amsfonts}
\usepackage{graphicx}
\usepackage{textcomp}
\usepackage{xcolor}
\usepackage{bm} 
\usepackage{amsmath}
\usepackage{algorithm}
\usepackage{algpseudocode}
\def\BibTeX{{\rm B\kern-.05em{\sc i\kern-.025em b}\kern-.08em
    T\kern-.1667em\lower.7ex\hbox{E}\kern-.125emX}}
\begin{document}

\title{A Hybrid Dynamic Model for Predicting Human Cognition and Reliance during Automated Driving
\thanks{This material is based upon work supported by the National Science Foundation under Award No. 2145827. Any opinions, findings, and conclusions or recommendations expressed in this material are those of the authors and do not necessarily reflect the views of the National Science Foundation.}
}

\author{\IEEEauthorblockN{Sibibalan Jeevanandam, Neera Jain}
\IEEEauthorblockA{\textit{School of Mechanical Engineering} \\
\textit{Purdue University}\\
West Lafayette, USA \\
sjeevana@purdue.edu, neerajain@purdue.edu}
}

\maketitle

\begin{abstract}
We propose a simple (12 parameter) hybrid dynamic model that simultaneously captures the continuous-valued dynamics of three human cognitive states---trust, perceived risk, and mental workload---as well as discrete transitions in reliance on the automation. The discrete-time dynamic evolution of each cognitive state is modeled using a first-order affine difference equation. Reliance is defined as a single discrete-valued state, whose evolution at each time step depends on the cognitive states satisfying certain threshold conditions. Using data collected from 16 participants, we estimate participant-specific model parameters based on their reliance on the automation and intermittently self-reported cognitive states during a continuous drive in a vehicle simulator. The model can be estimated using a single user's trajectory data (e.g. 8 minutes of driving), making it suitable for online parameter adaptation methods. Our results show that the model fits the observed trajectories well for several participants, with their reliance behavior primarily influenced by trust, perceived risk, or both. Importantly, the model is interpretable, such that the variations in model parameters across participants provide insights into differences in the time scales over which cognitive states evolve, and how these states are influenced by task complexity. Implications on the design of human-centric vehicle automation design are discussed.
% We further analyze the variability in model parameters across participants who are represented well by the model structure. 
\end{abstract}

% \begin{IEEEkeywords}

% \end{IEEEkeywords}

\section{Introduction}
\input{01_introduction_Jeevanandam_Jain_IEEE_ITSC_2025}

\section{Hybrid Model Structure}\label{sec:model}
\input{02_model_structure_Jeevanandam_Jain_IEEE_ITSC_2025}

\section{Experiment Design}\label{sec:experiment}
\input{03_human_experiment_Jeevanandam_Jain_IEEE_ITSC_2025}

\section{Parameter Identification}\label{sec:sys_id}
\input{04_parameter_identification_Jeevanandam_Jain_IEEE_ITSC_2025}

\section{Discussion}\label{sec:discussion}
\input{05_discussion_Jeevanandam_Jain_IEEE_ITSC_2025}

\section{Conclusion}\label{sec:conclusion}
\input{06_conclusion_Jeevanandam_Jain_IEEE_ITSC_2025}

\section*{Acknowledgment}

We would like to thank Prof. Brandon Pitts for providing valuable feedback on our experiment design, Xipeng Wang for developing the driving simulator software, and Tyler Hsieh for helping with experiment ideation.

\bibliographystyle{ieeetr}
\bibliography{references}

\end{document}

%% file: 01_introduction_Jeevanandam_Jain_IEEE_ITSC_2025.tex
Absent full automation in automated vehicles, effective human-automation interaction (HAI) necessitates  consideration of the human driver's cognitive state-space---including states such as the driver's trust in the automation. 
Computational models that capture the dynamics of multiple cognitive states and their relationship to human decision-making (e.g., reliance on the automation) can enable development of advanced autonomous systems aware of, and responsive to, the human driver. Several researchers have either examined cognitive factors such as trust \cite{stapel_-road_2022, sonoda_displaying_2017}, mental workload \cite{stapel_automated_2019}, and perceived risk \cite{stapel_-road_2022, perello-march_how_2024, li_no_2019}, or investigated their influence on takeover behavior such as reliance \cite{xiao_what_2025, abbasi_investigation_2024} and post-takeover performance \cite{melnicuk_effect_2021} in automated driving contexts. Given the dominant influence of trust on reliance \cite{lee_trust_1994, xiao_what_2025}, many researchers have developed computational models that capture how trust, or other cognitive states, directly influence reliance behavior---Table \ref{tab:lit_review} provides a summary of recent modeling efforts. See \cite{rodriguez_rodriguez_review_2023} for a detailed review.
\input{literature_review}
The first class of quantitative models include linear time-invariant (LTI) state-space and auto-regressive models \cite{hu_computational_2019,azevedo-sa_real-time_2021,hu_dynamic_2024}. These approaches, however, do not model reliance as inherently discrete-valued; rather, reliance is converted into a continuous-valued quantity by averaging it over (i) user trajectories in the sample population \cite{hu_computational_2019}, or (ii) time windows (e.g., percentage of time automation is used \cite{azevedo-sa_real-time_2021,hu_dynamic_2024}). The second class include variations of Markov Decision Process (MDP) models \cite{akash_toward_2020, williams_computational_2023} which instead discretize the cognitive states and represent their evolution using finite-state transitions. 

However, human cognitive states can exist on a continuum \cite{muir_trust_1996}, while reliance on the automation, by nature, is discrete-valued and often binary. Existing models that do account for the continuous- and discrete-valued nature of the cognitive states and reliance respectively, use Dynamic Bayesian Networks (DBN)\cite{xu_optimo_2015} or linear stochastic difference equations (SDE) \cite{ji_gao_extending_2006}. However, MDP, DBN, and SDE-type models are probabilistic, requiring significant amounts of data to estimate a large number of model parameters or probability distribution functions. Moreover, while researchers acknowledge the influence of multiple cognitive states---trust \cite{azevedo-sa_real-time_2021}, workload \cite{sato_automation_2020}, and risk perception \cite{li_no_2019}---on decision-making during automated driving, a comprehensive model that captures the dynamics of all three states is missing from the literature. 
% In summary, \emph{a computational model that accounts for the nature of the cognitive states (continuous-valued) and reliance (discrete-valued) identifiable with limited data is absent.}

To address these gaps, we propose a hybrid dynamic model structure that (1) accounts for the continuous-valued nature of multiple cognitive factors and the discrete-valued nature of reliance, (2) is interpretable and amenable to control-theoretic analysis and design, and (3) can be trained using a single participant's trajectory, enabling personalization. To train the model, we first design an experiment to elicit changes in human drivers' cognitive states and reliance behavior while interacting with an SAE level 3 automated vehicle in a simulated environment. We collect data from 16 participants through a human user study, identify participant-specific models, and interpret the resulting model parameters.

This paper is organized as follows. First, we describe the hybrid dynamic model structure and list simplifying assumptions based on cognitive psychology in Section \ref{sec:model}. In Section \ref{sec:experiment}, we describe our human experiment design and protocol. We discuss the parameter identification method and results in Section \ref{sec:sys_id}, analyze and interpret the estimated parameter values in Section \ref{sec:discussion}, and conclude the paper in Section \ref{sec:conclusion}.

%% file: literature_review.tex
\begin{table*}[ht!]
    \centering
    \caption{Computational models that capture the relationship between human cognitive states and reliance on automation.}
    \begin{tabular}{c c c c c c}
    \multicolumn{4}{c}{} & \multicolumn{2}{c}{\textbf{Continuous-/Discrete-Valued}}\\
    \hline
    \textbf{Paper}  & \textbf{Driving} & \textbf{Cognitive State(s)} & \textbf{Modeling Framework} & \textbf{Cognitive States} & \textbf{Reliance}\\
    \hline
    % Lee and Moray (1994) \cite{lee_trust_1994} & & Trust, Self-Confidence & ARMAV & Continuous & Continuous\\
    Hu et al. (2019) \cite{hu_computational_2019} & & Trust & LTI State-Space & Continuous & Continuous\\
    Azevedo-Sa et al. (2021) \cite{azevedo-sa_real-time_2021} & \checkmark & Trust & LTI State-Space\ & Continuous & Continuous\\
    Hu and Huang et al. (2024) \cite{hu_dynamic_2024} & \checkmark & Trust, Risk Perception & LTI State-Space & Continuous & Continuous \\
    Akash et al. (2020) \cite{akash_toward_2020} & \checkmark & Trust, Workload & Partially Observable MDP & Discrete & Discrete \\
    Williams et al. (2023)  \cite{williams_computational_2023} &  & Trust, Self-Confidence & Partially Observable MDP & Discrete & Discrete\\
    Xu and Dudek (2015) \cite{xu_optimo_2015} & &  Trust & Dynamic Bayesian Network & Continuous & Discrete\\
    Gao and Lee (2006) \cite{ji_gao_extending_2006} & & Trust, Self-Confidence & Linear SDE & Continuous & Discrete\\
    \textbf{This work} & \checkmark & Trust, Risk Perception, Workload & Piecewise Affine & Continuous & Discrete \\
    \hline
    \end{tabular}
    \label{tab:lit_review}
\end{table*}

%% file: 02_model_structure_Jeevanandam_Jain_IEEE_ITSC_2025.tex
Consider a discrete-time piece-wise affine (PWA) model with the continuous-valued state $x \in \mathbb{R}^n$ and the discrete-valued state $q \in  \{0,1,...,r\}\subseteq \mathbb{Z}$. The continuous-valued state evolves according to
\begin{equation}\label{eq:PWA_cont}
    x(k+1) = A_{q(k)}x(k) + B_{q(k)}u(k) + c_{q(k)} \enspace ,
\end{equation}
where $u\in \mathbb{R}^m$ denotes the exogenous input, and $k=0,1,\cdots,N$ denotes the discrete time index (with sampling time $T_s$). Additionally, $A_{q(k)} \in \mathbb{R}^{n \times n}$ and $B_{q(k)}\in \mathbb{R}^{n \times m}$ denote the state and input matrices respectively, while $c_{q(k)}\in \mathbb{R}^{n}$ accounts for non-stationarity in $x$. Subscript $q(k)$ indicates a dependence of the model parameters on the discrete-valued state $q$ at time index $k$. The evolution of $q$ is assumed to be dependent on the continuous state $x$ and is given by
\begin{equation}\label{eq:PWA_disc}
    q(k+1)=j\iff x(k+1) \in \mathcal{S}_j, \;j=0,\cdots,r \enspace ,
\end{equation}
where the sets $\mathcal{S}_j\subseteq\mathbb{R}^n$ partition the domain of $x$ such that 

\begin{equation*}
    \bigcup_{j=0}^r \mathcal{S}_j = \mathbb{R}^n, \; \bigcap_{j=0}^r \mathcal{S}_j=\phi.
\end{equation*}

We adopt this structure for modeling continuous-valued human cognitive states and discrete-valued reliance on the automation during automated driving. First, we define the continuous-valued state vector to include trust in the automation ($T$), risk perceived by the user ($R$), and mental workload ($W$), such that $x=\begin{bmatrix}
    T & R & W
\end{bmatrix}^T \in \mathbb{R}^3$. The discrete-valued state $q\in \{0,1\}$ denotes the user's reliance on the automation, where $q=1$ indicates that the automation is active, and $q=0$ indicates that it is inactive. Although reliance could be defined in terms of multiple discrete modes---such as lane keeping assist and adaptive cruise control being independently activated or deactivated---here we are interested in the driver's overall engagement with the automation. Hence, reliance is treated as a binary construct. The exogenous input $u \in \mathbb{R}^m$ can generally include controllable, automation-related factors such as transparency or measurable disturbances in the form of driving conditions (e.g. construction on the road or reduced visibility). The latter can be abstracted as a single exogeneous disturbance $d \in \mathbb{R}$ defined as the task complexity associated with the road environment. 

In this work, we model the evolution of the cognitive states ($x$) and reliance ($q$) during changes in task complexity ($d$). For the automated driving context, $u(k)$ in Eqn. \eqref{eq:PWA_cont} is replaced by $d(k)$ to emphasize that the input is a disturbance and not controllable. To reduce the number of model parameters, the PWA model (described by Eqns. \ref{eq:PWA_cont} and \ref{eq:PWA_disc}) is simplified based on the following assumptions.
\begin{itemize}
    \item The continuous state matrix $A$ is diagonal. Previous work suggests that while states such as trust and workload can be coupled, this coupling is weak \cite{akash_reimagining_2020}.
    \item The continuous state evolution is independent of the discrete state. In other words, $A$, $B$, and $c$ are not dependent on $q(k)$, such that $A_0=A_1$, $B_0=B_1$, and $c_0=c_1$.
    \item The human driver uses the automation when one or more of the following three conditions is met: trust is high \cite{abbasi_investigation_2024}, risk perceived is low \cite{abbasi_investigation_2024}, or mental workload is high \cite{sato_automation_2020}, relative to individually identified thresholds.
    \end{itemize}
    Consequently, we define the set
    \begin{equation*}
        \mathcal{S}_1=\left\{\begin{bmatrix}
            T\\R\\W
        \end{bmatrix}\in\mathbb{R}^3: T> \theta_T \text{ or }R<\theta_R\text{ or }W>\theta_W\right\}
    \end{equation*}
    and $\mathcal{S}_0=\mathcal{S}'_1$, i.e., $\mathcal{S}_0$ is the complement of $\mathcal{S}_1$. The parameters $\theta_T$, $\theta_R$, $\theta_W \in \mathbb{R}$ represent the aforementioned thresholds and are assumed to be constant for a given individual. Under these assumptions, the dynamic model for human cognitive states and reliance is given by
\begin{align}
    T(k+1) &= a_TT(k) + b_Td(k) + c_T \label{eq:T_update} \enspace ,\\
    R(k+1) &= a_R R(k) + b_R d(k) + c_R \label{eq:R_update} \enspace ,\\
    W(k+1) &= a_W W(k) + b_W d(k) + c_W \label{eq:W_update} \enspace ,
\end{align}
\begin{align}
\text{q}(k+1) = 
\begin{cases}
1, & \text{if } T(k+1) > \theta_T \\
  & \quad \text{or } R(k+1) < \theta_R \\
  & \quad \text{or } W(k+1) > \theta_W \\
0, & \text{otherwise} \enspace .
\end{cases}
\label{eq:reliance}
\end{align}

%% file: 03_human_experiment_Jeevanandam_Jain_IEEE_ITSC_2025.tex
\begin{figure*}[t]
    \centering
    \begin{minipage}[t]{0.68\textwidth}
        \centering
        \includegraphics[width=\linewidth]{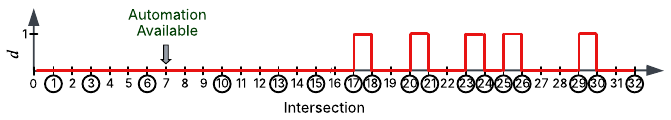}
        \caption{Binary signal representing task complexity during the main drive.}
        \label{fig:input_signal}
    \end{minipage}
    \hfill
    \begin{minipage}[t]{0.28\textwidth}
        \centering
        \includegraphics[width=\linewidth]{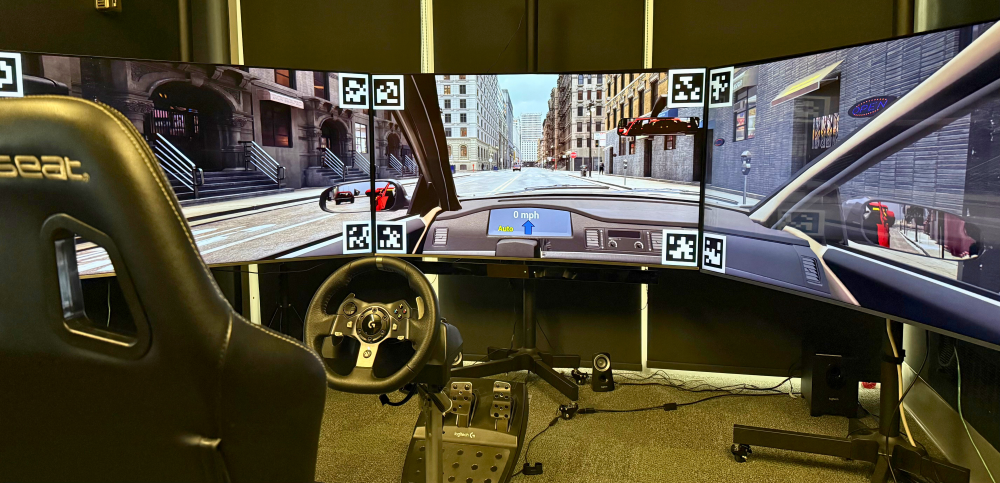}
        \caption{A photo of the driving simulator used in this study.}
        \label{fig:simulator}
    \end{minipage}
\end{figure*}

% \begin{figure*}[ht!]
%     \centering
%     \includegraphics[width=\linewidth]{Figures/Experiment/Input_Signal.pdf}
%     \caption{Binary signal representing task complexity during the main drive.}
%     \label{fig:input_signal}
% \end{figure*}
\begin{figure*}[t]
    \centering
    \includegraphics[width=\linewidth]{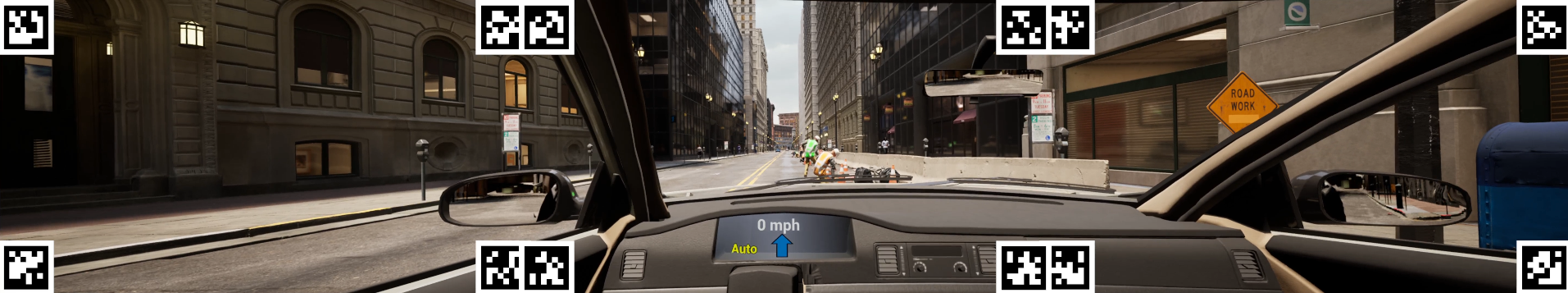}
    \caption{Participant's point-of-view when approaching a construction zone.}
    \label{fig:construction_pov}
\end{figure*}
% \begin{figure}[ht!]
%     \centering
%     \includegraphics[width=\linewidth]{Figures/Experiment/simulator.png}
%     \caption{A photo of the driving simulator used in this study.}
%     \label{fig:simulator}
% \end{figure}

We design an experiment to perturb the human driver's cognitive states ($x$) and consequently their reliance ($q$) on the automation via changes in task complexity ($d$). Each participant is tasked with navigating an urban environment along a pre-defined route in an ego-vehicle with SAE Level 3 automation during a single, continuous drive. During the drive, task complexity is manipulated in the form of a binary signal (shown in Figure \ref{fig:input_signal}) by placing construction zones along the route. Low complexity ($d=0$) corresponds to roads with low traffic density and pedestrians on the sidewalks, while high complexity ($d=1$) corresponds to the presence of construction zones with active (animated) human workers (Figure \ref{fig:construction_pov}), forcing vehicles to merge into the left lane. While the modeling framework supports a continuous-valued input, we manipulate it as a binary signal ($d \in {0,1}$) to simplify signal definition and achieve a low crest factor \cite{alma9932757301082}. The experiment is conducted in a custom-built medium-fidelity driving simulator (Figure \ref{fig:simulator}) consisting of a Logitech G920 steering wheel and pedal setup (used to control the ego-vehicle) and three Samsung 54.6" 4K Ultra HD Curved TVs which display the driving environment. The simulator software is developed using Unreal Engine 5. Furthermore, timestamps are recorded for relevant events, including driver engagement/disengagement of the automation or entering/exiting a construction zone. Finally, self-reports for cognitive factors (Table \ref{tab:self_reports}) are solicited at intersections (circled in Figure \ref{fig:input_signal}) by pausing the simulation at a red traffic light or a stop sign. Participants respond on a scale ranging from 0 (labeled ``Very Low") to 100 (labeled ``Very High") in increments of 5.

\begin{figure*}[t]
    \centering
    \includegraphics[width=\linewidth]{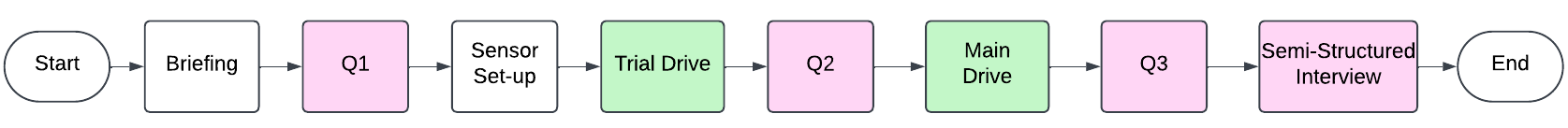}
    \caption{Flowchart describing the experimental procedure. Pink boxes depict steps involving subjective data collection, while green boxes involve use of the driving simulator.}
    \label{fig:procedure}
\end{figure*}

\subsection{Procedure}
Figure \ref{fig:procedure} provides an overview of the experimental procedure. First, participants are briefed about the task and informed that the ego-vehicle ``can monitor its surroundings and control both its speed and steering at the same time." They are instructed to pay attention and always be ready to take control of the vehicle to avoid any accidents. Participants are also informed of a \$5 bonus for completing the trip without any accidents or traffic violations. This incentive is intended to encourage careful driving and ensure that participants take the consequences of crashing seriously within the simulated environment. Nevertheless, all participants receive the bonus at the end of the experiment, regardless of their performance. Throughout the experiment, the automation is 100\% reliable and capable of completing the entire route without errors. However, participants are not informed of this fact. Participants complete a questionnaire (see Q1 in Figure \ref{fig:procedure}) designed to capture dispositional factors such as automation bias, driving self-efficacy, driving-related risk propensity, and driving style. Next, multiple sensors are attached to the participant, including a Polar H10 heart rate monitor, a Shimmer3 GSR+ unit, a NIRSport2 fNIRS (functional near-infrared spectroscopy) cap by NIRx, and Neon eye-tracking glasses by Pupil Labs. Note that data collected using these sensors are not analyzed in this paper. 

Following sensor set up, participants receive instructions regarding the simulator controls and the self-reporting process. Participants are then given a trial drive to practice driving in the simulator---this also includes navigating through a construction zone for them to assess their ability to navigate it in the simulator. At the end of the trial drive, participants are asked to rate their ability (Q2) to complete driving tasks in the simulator, such as making turns, engaging/disengaging the automation, and navigating through the construction zone. They are also informed that the automation requires time to get acclimated to the driving conditions and will be unavailable at the start of the main drive. This period is intended at capturing participants' baseline responses in different sensing modalities during manual driving. Finally, participants are told that once the automation prompts them about its availability, they may engage or disengage it at their discretion. Upon completing the main drive, participants complete another questionnaire (Q3) to report demographic information. The experiment concludes with a short semi-structured interview.

% \subsection{Participants}
The study was approved by Purdue University's Institutional Review Board. Upon obtaining informed consent, 16 participants (4 males and 12 females), aged between 19 and 31 (Mean=22.88, SD=3.74) with a valid US driver's license participated in the study. Participants were compensated at a rate of \$5 per 15 minutes, with the experiment lasting approximately one hour on average.

\subsection{Data Processing}
First, we construct the time-series $d(k)$ and $q(k)$ for each participant using the timestamps recorded for entering/exiting a construction zone and engaging/disengaging the automation respectively, for a choice of sampling time ($T_s=1$). Next, we round the timestamps associated with self-reports to the nearest discrete-time instants ($kT_s$), and represent the set of discrete-time indices ($k$) at which self-reports are available as ${K}_{SR}$. Additionally, without loss of generality, the self-reported cognitive states are scaled down by a factor of 100 so that they are defined on a scale of 0 to 1. Finally, to mitigate the effects of initial transients in the cognitive factors soon after the automation is made available, we use data starting from intersection 15 (see Figure \ref{fig:input_signal}) for the identification process. Thus, for each participant, the following data is available: 
\begin{itemize}
    \item task complexity, $d(k)\; \forall k=0,\cdots,N$,
    \item cognitive states self-reported by the participant, $\begin{bmatrix}T(k)&R(k)&W(k)\end{bmatrix}^T \; \forall k \in {K}_{SR}$, and
    \item reliance, $q(k) \;\forall k=0,\cdots,N$.
\end{itemize}
Here, $k=0$ denotes the discrete-time index corresponding to the ego-vehicle arriving at intersection 15 in Figure \ref{fig:input_signal}, and $k=N$ denotes the end of the main drive (intersection 32).

\input{self_reports}

%% file: self_reports.tex
\begin{table}[t]
    \centering
    \caption{Prompts for soliciting self-reported cognitive states.}
    \label{tab:self_reports}
    \begin{tabular}{l l}
        \hline
        \textbf{State} & \textbf{Prompt} \\ \hline
        $T$ & What is your current level of trust in the automation? \\
        $W$ & What is your current level of mental workload \\
        $R$ & What is the current risk of an accident? \\ 
        \hline
    \end{tabular}
\end{table}

%% file: 04_parameter_identification_Jeevanandam_Jain_IEEE_ITSC_2025.tex
We identify participant-specific model parameters using each participant's entire trajectory ($k=0,\cdots,N$) to evaluate whether our proposed hybrid model structure adequately fits each participant's data. The set of parameters to be identified include the model parameters for the continuous states ($a_T,b_T,c_T,a_R,b_R,c_R,a_W,b_W,c_W$) and the thresholds ($\theta_T,\theta_R,\theta_W$).

\subsection{Identification Procedure}
\input{algorithm_continuous}
Since the cognitive state dynamics are assumed to be uncoupled, the model parameters for trust ($a_T, b_T, c_T$ in Eqn. \ref{eq:T_update}) can be estimated independently of the perceived risk and workload models (Eqns. \ref{eq:R_update} and \ref{eq:W_update}). Moreover, the independence of the continuous state evolution from the discrete state allows us to estimate the thresholds ($\theta_{T},\theta_{R},\theta_{W}$) independently of the continuous state model parameters. Consequently, we first estimate the continuous state model parameters (Algorithm \ref{alg:continuous}), followed by the discrete state model parameters (Algorithm \ref{alg:seq}). While Algorithm \ref{alg:continuous} is described in the context of estimating the trust model parameters ($a_T, b_T, c_T$), the same procedure is applied to estimate the perceived risk ($a_R, b_R, c_R$) and workload parameters ($a_W, b_W, c_W$). The optimization problem in Algorithm \ref{alg:continuous} is solved in MATLAB using \emph{fminsearch} \cite{OptimizationToolbox}, which uses the Nelder-Mead simplex algorithm \cite{lagarias_convergence_1998}. The optimization in Algorithm \ref{alg:seq} is solved using a genetic algorithm in MATLAB's \emph{Global Optimization Toolbox} \cite{GeneticAlgorithm}.
\input{algorithm_discrete}

 To test each identified model, we simulate the trajectory from the initial condition ${x}(0)$ using task complexity $d(k)$ and compare the continuous state and discrete state prediction to the true trajectory. To quantify the model fit, we compute the root mean squared error (RMSE) for continuous state predictions. For the trust state $T$, this is given by
\begin{equation}
    \text{RMSE}_T = \sqrt{\frac{1}{card({K}_{SR})} \sum_{k\in {K}_{SR}} (T(k) - \hat{T}(k))^2},
\end{equation}
where $card(.)$ denotes set cardinality. Additionally, we also evaluate discrete state prediction accuracy as defined in Algorithm \ref{alg:seq}. We consider the model to represent a participant well if RMSE lie within 10\% of the range of self-reports (RMSE $\leq 0.1$) for all cognitive states, and discrete state prediction accuracy is greater than 80\%. Note that while predicting the cognitive states, an error of up to 0.025 may result solely by virtue of the discretization of the self-report scale (increments of 5 on the 100 point scale).
% self-reporting the cognitive states in increments of 0.05 (5 scaled down by 100).

\subsection{Results}
Table \ref{tab:results_accuracy} summarizes the model fit, in terms of RMSE and accuracy, for each participant-specific model. It also includes the active thresholds ($\theta_{act}$) which are crossed by their respective cognitive states at least once 
during the predicted trajectory, resulting in a change in the discrete state. A missing active threshold (--) indicates that none of the cognitive states crossed their corresponding thresholds during prediction. Notably, most participants' reliance on the automation is driven by their trust, perceived risk, or a combination of the two. Mental workload influences reliance for only one participant (01). Note that this is likely to be a consequence of the absence of a secondary task in our experiment---monitoring the automation alone may not induce a sufficiently high workload to influence reliance behavior. 

\begin{figure}[ht!]
    \centering
    \includegraphics[width=\linewidth]{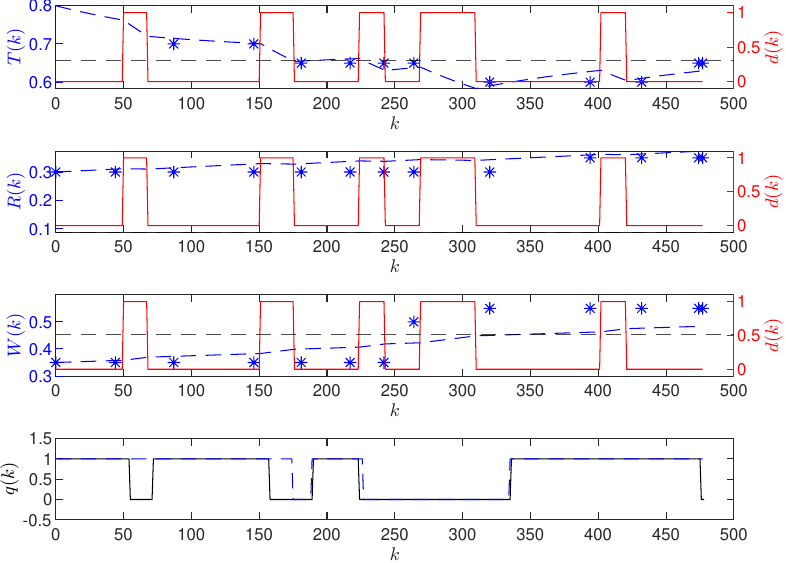}
    \caption{State trajectory for Participant 01. The blue dashed lines show the simulated trajectory generated by the identified model, while the black dashed lines indicate the estimated active thresholds. The blue asterisks denote the cognitive states self-reported by the participant. Fit:  RMSE$_T=0.0177$, RMSE$_R=0.0269$, RMSE$_W=0.0644$, Acc.(\%)=91.21.}
    \label{fig:P001_full}
\end{figure}

\begin{figure}[ht!]
    \centering
    \includegraphics[width=\linewidth]{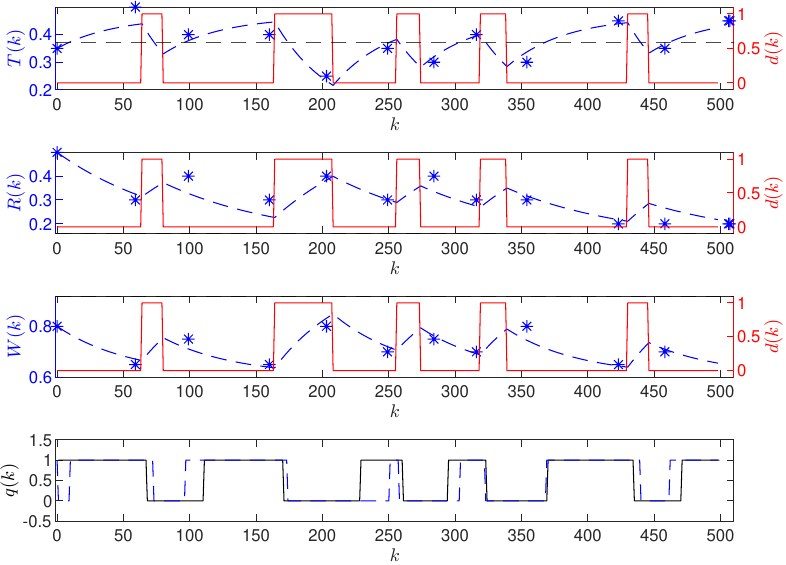}
    \caption{State trajectory for Participant 03. Fit:  RMSE$_T=0.0298$, RMSE$_R=0.0453$, RMSE$_W=0.0253$, Acc.(\%)=83.77.}
    \label{fig:P003_full}
\end{figure}

\begin{figure}[ht!]
    \centering
    \includegraphics[width=\linewidth]{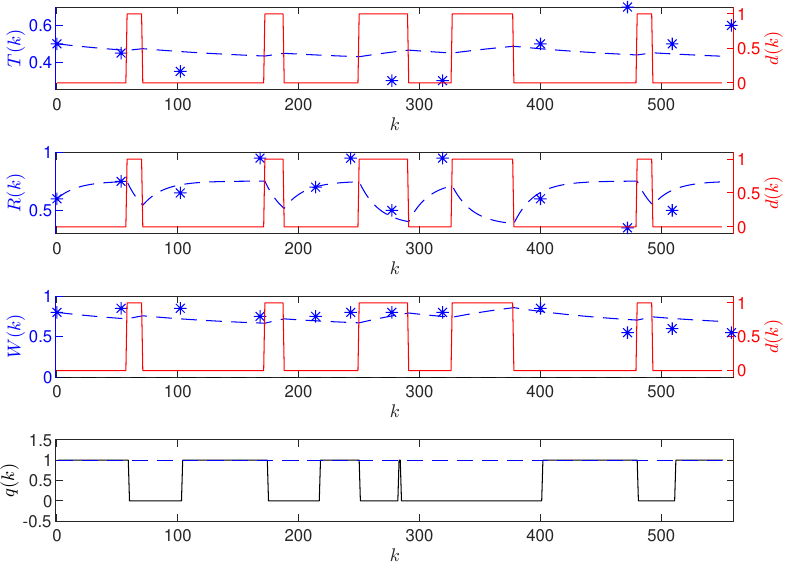}
    \caption{State trajectory for Participant 04. None of the thresholds describe the reliance behavior. Fit:  RMSE$_T=0.1526$, RMSE$_R=0.1876$, RMSE$_W=0.1026$, Acc.(\%)=51.54.}
    \label{fig:P004_full}
\end{figure}

For 9 of 16 participants, the identified models fit with all RMSE $\leq 0.1$. Moreover, the discrete state prediction accuracy is more than 80\% for 13 of 16 participants. The state trajectories are illustrated for three participants---two with good model fits (Figures \ref{fig:P001_full} and \ref{fig:P003_full}), and one with a poor fit (Figure \ref{fig:P004_full}).
\input{results_accuracy}

\begin{figure*}[ht!]
    \centering
    % First figure
    \begin{minipage}[t]{0.29\textwidth} % Use 't' for top alignment
        \centering
        \includegraphics[width=\textwidth]{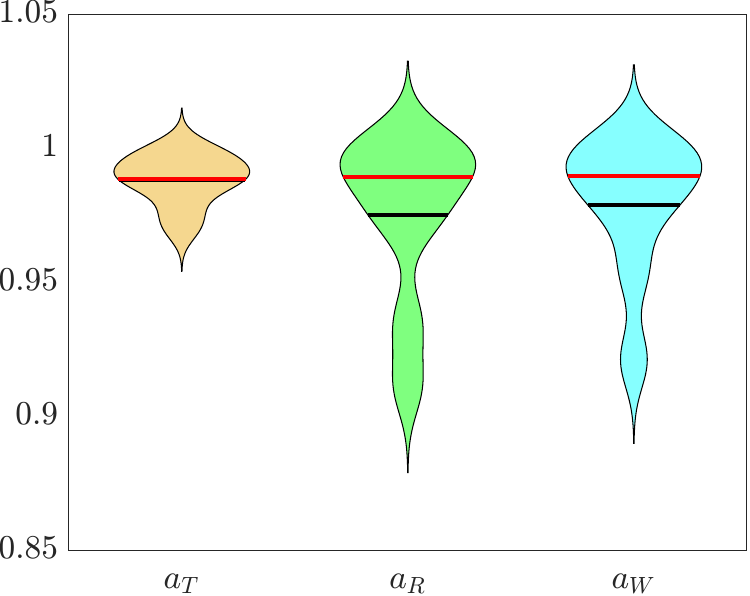}
        \caption{Violin plots depicting the distributions of $a_T$, $a_R$, and $a_W$. The red line denotes the mean value; the black line denotes the median.}
        \label{fig:dist_a}
    \end{minipage}
    \hspace{0.03\textwidth} % Adjust horizontal space
    % Second figure
    \begin{minipage}[t]{0.3\textwidth} % Use 't' for top alignment
        \centering
        \includegraphics[width=\textwidth]{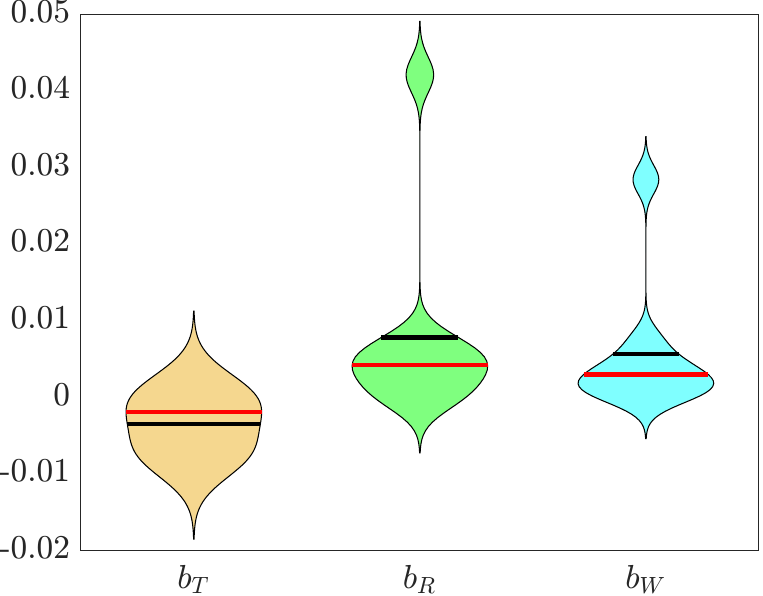}
        \caption{Violin plots depicting the distributions of $b_T$, $b_R$, and $b_W$.}
        \label{fig:dist_b}
    \end{minipage}
    \hspace{0.03\textwidth}
    % Third figure
    \begin{minipage}[t]{0.3\textwidth} % Use 't' for top alignment
        \centering
        \includegraphics[width=\textwidth]{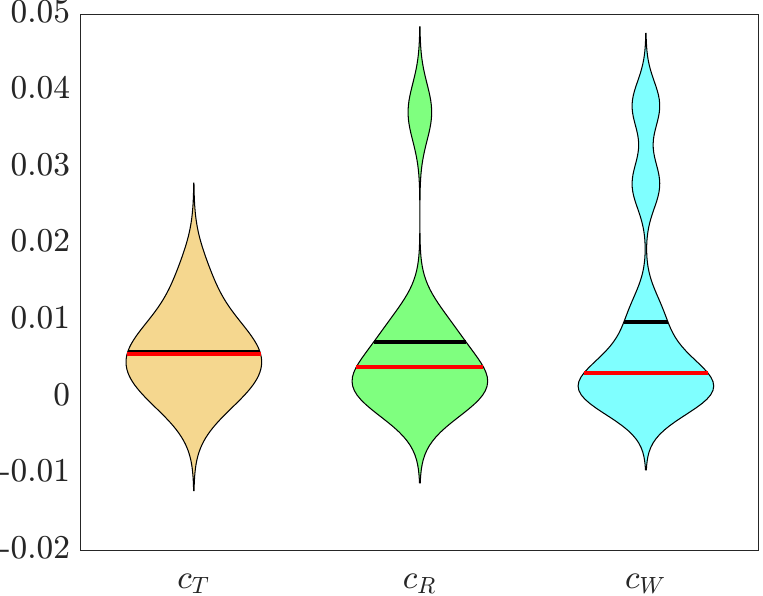}
        \caption{Violin plots depicting the distributions of $c_T$, $c_R$, and $c_W$.}
        \label{fig:dist_c}
    \end{minipage}
\end{figure*}

For the models that predict the continuous state well (all RMSE $\leq 0.1$), we also estimate the distribution of the continuous state model parameters ($a,b,c$) to understand how the model parameters vary between participants. We compute the kernel density estimate $\hat{f}$ for a given parameter $y$ as
\begin{equation}
\hat{f}(y) = \frac{1}{n_p h \sigma} \cdot \frac{1}{\sqrt{2\pi}} \sum_{i=1}^{n_p} \exp\left( \frac{-(y - y_i)^2}{2 h^2 \sigma^2} \right) ,
\end{equation}
\noindent where, $\sigma$ denotes the standard deviation of parameters $y_i$, $n_p$ denotes the number of parameter samples, and $h$ denotes the bandwidth chosen using Silverman's rule \cite{silverman_density_2018}. The estimated distributions are visualized using violin plots shown in Figures \ref{fig:dist_a}, \ref{fig:dist_b}, and \ref{fig:dist_c}; the contours of the plots represent the estimated distributions. Note that given the nonlinear cost formulation in Algorithm \ref{alg:continuous}, estimating theoretical uncertainty bounds on the model parameters is nontrivial and therefore not explored here.
 
\subsection{Predictive Capability on Unseen Data}
\input{validation_accuracy}

\begin{figure}[t]
    \centering
    \includegraphics[width=\linewidth]{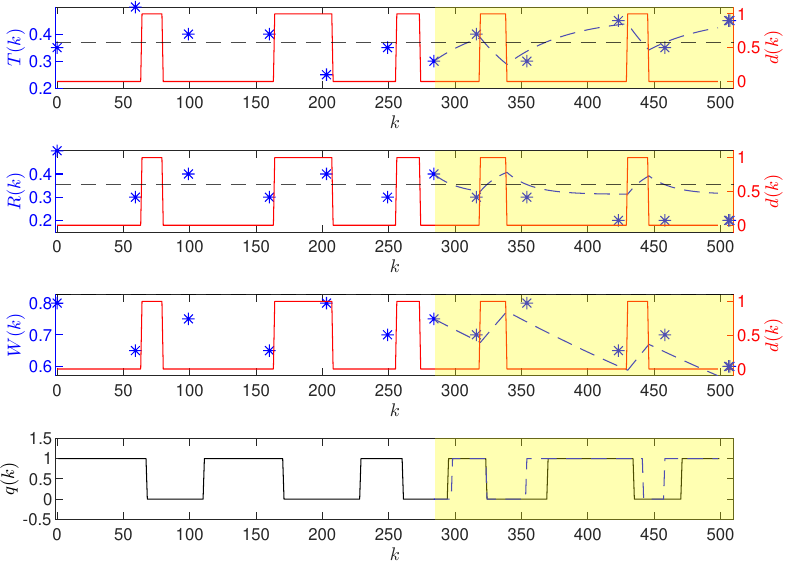}
    \caption{State trajectory for Participant 03 when model parameters are estimated using the first half of their trajectory and used for prediction in the second half (highlighted in yellow).}
    \label{fig:P003_val}
\end{figure}

For participants whose trajectory is well described by the model structure, we also investigate the model's capability to predict a part of a trajectory that is not included during parameter estimation. Thus, for each participant in Table \ref{tab:results_accuracy} with all RMSE $\leq 0.1$ and discrete state accuracy greater than 80\%, we divide the trajectory into two halves based on the number of self-reports. We then identify model parameters using the first half of the trajectory, and validate the identified model on the remaining half. Table \ref{tab:validation_accuracy} summarizes results from testing the identified models. The simulated state trajectory is illustrated for one participant with a good model validation accuracy in Figure \ref{fig:P003_val}.

\begin{figure}[t]
    \centering    \includegraphics[width=\linewidth]{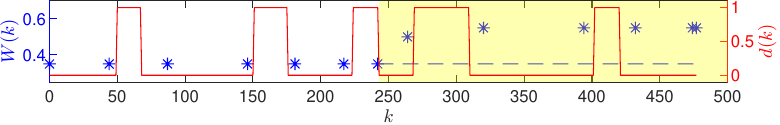}
    \caption{Workload trajectory for Participant 01. The self-reported data lacks sufficient excitation in the first half.}
    \label{fig:01_workload}
\end{figure}

%% file: algorithm_continuous.tex
\begin{algorithm}[t]
\caption{Identification of continuous state model parameters using intermittent self-report data.}\label{alg:continuous}
\begin{algorithmic}
% \State \textbf{Initialize:}
\State $E_{T} = [\;]$  \Comment{Prediction error list}
\State $\hat{T}(0) = T(0)$  \Comment{Set initial value to measured value}\\
\State $k=1$
\While{$k \leq N$}       \Comment{Predict trajectory}
\State $\hat{T}(k) = a_T\hat{T}(k-1) + b_Td(k-1) + c_T$\Comment{Prediction}\\
\If{$k \in K_{SR}$} \Comment{If self-report is available}
\State $e_{T} = \hat{T}(k) - T(k)$     \Comment{Prediction error}
\State $E_{T} = [E_{T}; \;e_{T}]$ \Comment{Add error to list}
\State $\hat{T}(k) = T(k)$  \Comment{Reset regressor to measured value}\\
\EndIf
\State $k+=1$
\EndWhile\\

\State \textbf{Optimization problem for identification:}
\State $[a^*_T,b^*_T,c^*_T]=\underset{a_T,b_T,c_T}{\mathrm{argmin}} \; ||E_T||_2 $
\State s.t. $|a_T|<1$ \Comment{Only identify stable models}
\end{algorithmic}
\end{algorithm}

%% file: algorithm_discrete.tex
\begin{algorithm}[t]
\caption{Identification of discrete state model parameters given the continuous state parameters ($a^*,b^*,c^*$).}\label{alg:seq}
\begin{algorithmic}

% \State \textbf{Initialize:}
\State $\hat{{Q}}= [\;]$  \Comment{Reliance prediction list}
\State $Q=[q(0),\cdots,q(N)]$ \Comment{Ground truth}
\State $\hat{q}(0) = q(0)$ \Comment{Set initial values}
\State $\hat{T}(0) = T(0), \; \hat{R}(0) = R(0), \;\hat{W}(0) = W(0)$ \\
\State $k=1$

\While{$k\leq N$}       \Comment{Predict trajectory}
\State $\hat{T}(k) = a^*_T\hat{T}(k-1) + b^*_Td(k-1) + c^*_T$
\State $\hat{R}(k) = a^*_R\hat{R}(k-1) + b^*_Rd(k-1) + c^*_R$
\State $\hat{W}(k) = a^*_W\hat{W}(k-1) + b^*_Wd(k-1) + c^*_W$
\\
\State $\hat{q}(k)=\begin{cases}
    1, & \hat{T}(k)>\theta_T \text{ or }\hat{R}(k)<\theta_R\text{ or }\hat{W}(k)>\theta_W\\
    0, & \text{otherwise}
\end{cases}$

% \State $\hat{q}(k) = \begin{cases}
% 1, & \text{if } \hat{T}(k) \geq T_{\text{th}} \\
%   & \quad \text{or } \hat{R}(k) \leq R_{\text{th}} \\
%   & \quad \text{or } \hat{W}(k) \geq W_{\text{th}} \\
% 0, & \text{otherwise}
% \end{cases}$ \Comment{Discrete-state}
\State $\hat{{Q}} = [\hat{{Q}};\;\hat{q}(k)]$
\EndWhile\\

\State \textbf{Compute accuracy:} \Comment{$card(.)$ denotes cardinality}
\State $TP=card(\{k:Q(k)=1, \hat{Q}(k)=1\})$ \Comment{True positives}
\State $TN=card(\{k:Q(k)=0, \hat{Q}(k)=0\})$ \Comment{True negatives}
\State $FN=card(\{k:Q(k)=1, \hat{Q}(k)=0\})$ \Comment{False negatives}
\State $TN=card(\{k:Q(k)=0, \hat{Q}(k)=1\})$ \Comment{False positives}
\State $\text{Accuracy} = (TP + TN)/(TP+TN+FP+FN)$ \\
\State \textbf{Optimization problem for thresholds:}
\State $[\theta^*_T,\theta^*_R,\theta^*_W]=\underset{\theta_T,\theta_R,\theta_W}{\mathrm{argmax}} \; \text{Accuracy}$

\end{algorithmic}
\end{algorithm}

%% file: results_accuracy.tex
\begin{table}[t]
    \centering
    \caption{Model fit for participant-specific models.}
    \begin{tabular}{c c c c c c}
    \hline
        \textbf{ID} & $\bm{\text{RMSE}_T}$ & $\bm{\text{RMSE}_R}$ & $\bm{\text{RMSE}_W}$ & $\bm{Acc.}(\%)$ & $\bm{\theta_{act}}$ \\
    \hline
   01 & 0.0177 & 0.0269 & 0.0644 & 91.21 & T,W\\
    02 & 0.0296 & 0.0725 & 0.0388 & 89.80 & --\\
    03 & 0.0298 & 0.0453 & 0.0253 & 83.77 & T\\
    04 & 0.1526 & 0.1876 & 0.1026 & 51.54 & -- \\
    05 & 0.0618 & 0.0847 & 0.0509 & 61.24 & T\\
    06 & 0.0743 & 0.0253 & 0.0456 & 89.17 & T\\
    07 & 0.1053 & 0.1199 & 0.1299 & 80.45 & T,R\\ 
    08 & 0.1415 & 0.0863 & 0.0759 & 82.46 & T\\ 
    09 & 0.0662 & 0.0707 & 0.0606 & 86.61 & R\\ 
    10 & 0.2002 & 0.2930 & 0.1082 & 80.58 & T,R\\
    11 & 0.0002 & 0.1204 & 0.0868 & 100 & --\\
    12 & 0.0594 & 0.0688 & 0.0345 & 83.72 & R\\
    13 & 0.0533 & 0.0669 & 0.0246 & 59.47 & R\\ 
    14 & 0.1165 & 0.0927 & 0.0872 & 84.01 & T \\
    15 & 0.0832 & 0.0000 & 0.0325 & 88.53 & T\\ 
    16 & 0.0289 & 0.2143 & 0.0144 & 100 & --\\

    \hline
    \end{tabular}\\
    \label{tab:results_accuracy}
    Note: Acc. denotes Accuracy.
\end{table}

%% file: validation_accuracy.tex
\begin{table}[t]
    \centering
    \caption{Testing accuracy on unseen data.}
    \begin{tabular}{c c c c c}
    \hline
        \textbf{ID} & $\bm{\text{RMSE}_T}$ & $\bm{\text{RMSE}_R}$ & $\bm{\text{RMSE}_W}$ & $\bm{\text{Acc.}}(\%)$ \\
    \hline
        01 & 0.1212 & 0.0408 & 0.1924 & 40.68\\
        02 & 0.0491 & 0.1047 & 0.0418 & 81.82 \\
        03 & 0.0240 & 0.1003 & 0.0479 & 81.86\\
        06 & 0.0588 & 0.0337 & 0.0329 & 100.00\\
        09 & 0.1704 & 0.1175 & 0.0907 & 67.11\\
        12 & 0.1233 & 0.1526 & 0.1477 & 36.07\\
        15 & 0.2199 & 0.0000 & 0.0584 & 87.56\\
    \hline
    \end{tabular}
    \label{tab:validation_accuracy}
\end{table}

%% file: 05_discussion_Jeevanandam_Jain_IEEE_ITSC_2025.tex
% Using data collected from 16 participants in an in-person human study, we demonstrate the ability to estimate model parameters specific to the user from limited data. Although the continuous-valued cognitive states are measured only intermittently, the participant-specific models predict all three cognitive states with a $RMSE \leq 0.1$ for more than half of the participants (see Table \ref{tab:results_accuracy}). The affine model structure for cognitive states mitigates overfitting through low model complexity, while still achieving strong prediction accuracy with limited data. Furthermore, the discrete state (reliance) prediction accuracy exceeds 80\% for 13 of 16 participants, validating the use of a hybrid dynamic model and describing the evolution reliance to be a cognitive state-dependent, threshold-based process. Furthermore, not all cognitive factors influenced reliance, evident from the active thresholds ($\theta_{act}$) in Table \ref{tab:results_accuracy}. 
 % Nevertheless, our model structure provides the flexibility to adapt to individuals with reliance behaviors influenced by any or all of the three cognitive states. 
\subsection{Implications for Human-Centric Vehicle Automation Design}
A key feature of our hybrid model structure lies in its threshold-based description of reliance. While individuals may self-report cognitive states across different ranges and interpret the extremes of the given scale (0-100) differently, the thresholds anchor these subjective measurements to make the predictions useful---particularly for adapting the vehicle's automation to individual users.  For instance, a predicted trust value of 0.6, in isolation, is insufficient to determine if an individual's trust is high. However, knowledge of their threshold (e.g. $\theta_T=0.7$) allows the automation to interpret their trust---the individual is 0.1 units away from engaging the automation. This can inform personalized interventions such as take-over requests (TORs). For example, consider individual X and individual Y with similar trust dynamics ($a_T,b_T,c_T$). Let X be a high-trusting individual with $T=0.9,\theta_T=0.3$, and Y be moderately trusting, with $T=0.5, \theta_T=0.4$. If the automation predicts that it may need the human driver to intervene in an upcoming complex driving scenario, individual X should be given a higher TOR lead time for them to safely retrieve control of the vehicle---this is because drivers excessively trustful of the automation are prone to poor takeover performance \cite{pan_how_2023}.

\subsection{Model Parameters and Inter-Participant Variation}
Another feature of the proposed model is that it is interpretable. For example, the parameter $a$ quantifies the inertia in the cognitive states, i.e. how much the future state depends on the current state. Figure \ref{fig:dist_a} shows the variation in $a_T$, $a_R$, and $a_W$ across participants. Among the three, the parameter variation is highest for risk ($a_R$) and lowest for trust ($a_T$). This aligns with previous work \cite{akash_toward_2020} which showed that trust tends to be a relatively steady trait when the automation is highly reliable. In fact, the estimated $a_T$ values are close to 1. On the contrary, risk perception is highly context-dependent---given an individual's disposition towards risk aversion, their perception of situational risks will be different---this necessitates personalized adaptation. Furthermore, although the mean values are similar, the median value for $a_T$ is the largest, while the median value for $a_R$ is the smallest. This suggests that, on average (in the median sense), participants' perceived risk changed the fastest, while trust changed the slowest.

The parameter $b$ captures the effect of task complexity on cognitive states. As shown in Figure \ref{fig:dist_b}, the average values (both mean and median) are negative for trust ($b_T$) and positive for risk ($b_R$) and workload ($b_W$). This aligns with intuition: during complex tasks--—such as navigating a construction zone—--participants’ trust tends to decrease, while their perceived risk and workload increase. 

% Additionally, the distributions of $b_T$ and $b_W$ are bimodal, suggesting the presence of two distinct user groups—--one that is significantly more affected by task complexity than the other. This insight provides a useful prior for model adaptation to new users; the parameters $b_R$ and $b_W$ parameters can be initialized near the two modes of the distributions.

Finally, the parameter $c$ captures non-stationary effects, specifically linear trends in the cognitive states over time. In Figure \ref{fig:dist_c}, the mean and median are positive for $c_T,c_R, \text{and }c_W$, indicating a steady increase (with time) in all three cognitive states. Finally, note that the range of the violin plot limits in Figure \ref{fig:dist_a} is 0.2, which is an order of magnitude larger than the range of 0.03 in Figures \ref{fig:dist_b} and \ref{fig:dist_c}, indicating substantially greater variation in parameter $a$ across participants. Note that the participants were predominantly female (12 out of 16), limiting the generalizability of these findings to the broader population. 

\subsection{Identification Using Limited Data}
For participants well-represented by our model structure, we fit the model parameters using the first half of their trajectory and evaluate their accuracy using the second half (see Table \ref{tab:validation_accuracy}). While 6 of the 7 participant models achieve RMSE $\leq0.2$ across all cognitive states, only one model (Participant 06) achieved all RMSE $\leq 0.1$. For some participants, the continuous state prediction is poor due to insufficient excitation in the self-reported cognitive states during the first half of the trajectory. For example, the model for Participant 01 (Figure \ref{fig:01_workload}) fails to predict changes in workload in the second half due to no changes in self-reported in workload in the first half. On the contrary, despite the limited data, the model achieves over 80\% accuracy in predicting the discrete state for 4 of the 7 participants tested. For example, for Participant 03 (see Figure \ref{fig:P003_val}), by estimating model parameters using 7 self-reports and approximately 5 minutes of reliance observed at 1Hz, the model is able to predict all three user-initiated takeovers in the second half of their interaction with the vehicle. This holds promise for personalized online adaptation of the thresholds for new users.

%% file: 06_conclusion_Jeevanandam_Jain_IEEE_ITSC_2025.tex
In this paper, we present a hybrid dynamic modeling approach for human cognitive states and reliance in the context of conditionally automated driving. Key highlights of our work include the following.
\begin{enumerate}
    \item The piecewise affine model simultaneously captures the continuous-valued dynamics of three human cognitive states as well as discrete transitions in reliance on the automation, departing from existing computational models by adapting to individuals whose reliance behaviors can be influenced by up to three cognitive states.
    \item The model is simple (with 3 parameters per cognitive state, and 3 thresholds for describing reliance) and can be estimated using a single user's trajectory data, lending itself to online parameter adaptation methods to capture participant-specific  behaviors.
    \item  The model is interpretable, such that the variations in model parameters across participants provide insights into differences in the time scales over which cognitive states evolve, and how these states are influenced by task complexity. 
\end{enumerate} 

These features move us closer to the practical realization of cognition-aware automated systems. In future work, we plan to evaluate the utility of non-disruptive sensing modalities---such as eye-tracking and heart rate monitoring---and additional context variables, such as distraction, for inferring cognitive states. We also aim to relax several assumptions made to the model to investigate if reliance-dependent evolution of the cognitive states ($A_0\neq A_1$) and coupling between cognitive states (non-zero off-diagonal elements in $A$) can capture participant behaviors not described well by the current model. Finally, we aim to adapt our model to account for the effect of controllable inputs, such as takeover requests and automation transparency, on reliance behavior, to enable its use for personalized driving interventions. 

%% file: references.bib
@article{pan_how_2023,
	title = {How does drivers’ trust in vehicle automation affect non-driving-related task engagement, vigilance, and initiative takeover performance after experiencing system failure?},
	volume = {98},
	issn = {1369-8478},
	url = {https://www.sciencedirect.com/science/article/pii/S1369847823001845},
	doi = {10.1016/j.trf.2023.09.001},
	urldate = {2025-07-15},
	journal = {Transportation Research Part F: Traffic Psychology and Behaviour},
	author = {Pan, Hengyan and Xu, Ke and Qin, Yang and Wang, Yonggang},
	month = oct,
	year = {2023},
	pages = {73--90},
}

@book{alma9932757301082,
author = {Ljung, Lennart.},
address = {Upper Saddle River, NJ},
booktitle = {System identification : theory for the user},
edition = {2nd ed.},
year = {1999},
isbn = {0136566952},
language = {eng},
lccn = {98018554},
publisher = {Prentice Hall PTR},
series = {Prentice Hall information and system sciences series},
title = {System identification : theory for the user },
}

@misc{OptimizationToolbox,
year = {2021},
author = {{The MathWorks Inc.}},
title = {Optimization Toolbox version: 9.2 ({R}2021b)},
publisher = {The MathWorks Inc.},
address = {Natick, Massachusetts, United States},
url = {https://www.mathworks.com}
}

@misc{GeneticAlgorithm,
year = {2021},
author = {{The MathWorks Inc.}},
title = {Global Optimization Toolbox version: 4.6 ({R}2021b)},
publisher = {The MathWorks Inc.},
address = {Natick, Massachusetts, United States},
url = {https://www.mathworks.com}
}

@article{rodriguez_rodriguez_review_2023,
	title = {A review of mathematical models of human trust in automation},
	volume = {4},
	issn = {2673-6195},
	url = {https://www.frontiersin.orghttps://www.frontiersin.org/journals/neuroergonomics/articles/10.3389/fnrgo.2023.1171403/full},
	doi = {10.3389/fnrgo.2023.1171403},
	language = {English},
	urldate = {2025-05-01},
	journal = {Frontiers in Neuroergonomics},
	author = {Rodriguez Rodriguez, Lucero and Bustamante Orellana, Carlos E. and Chiou, Erin K. and Huang, Lixiao and Cooke, Nancy and Kang, Yun},
	month = jun,
	year = {2023},
}

@article{xiao_what_2025,
	title = {What leads to reliance on automated vehicles? {An} inferential analysis of responses to variable {AV} performance},
	volume = {128},
	issn = {0003-6870},
	shorttitle = {What leads to reliance on automated vehicles?},
	url = {https://www.sciencedirect.com/science/article/pii/S000368702500047X},
	doi = {10.1016/j.apergo.2025.104511},
	urldate = {2025-05-01},
	journal = {Applied Ergonomics},
	author = {Xiao, Xizi and Ma, Xingjian and McDonald, Anthony D. and Mehta, Ranjana K.},
	month = oct,
	year = {2025},
	pages = {104511},
}

@book{silverman_density_2018,
	address = {New York},
	title = {Density {Estimation} for {Statistics} and {Data} {Analysis}},
	isbn = {978-1-315-14091-9},
	publisher = {Routledge},
	author = {Silverman, Bernard W.},
	month = feb,
	year = {2018},
	doi = {10.1201/9781315140919},
}

@article{lagarias_convergence_1998,
	title = {Convergence {Properties} of the {Nelder}--{Mead} {Simplex} {Method} in {Low} {Dimensions}},
	volume = {9},
	issn = {1052-6234},
	url = {https://epubs.siam.org/doi/10.1137/S1052623496303470},
	doi = {10.1137/S1052623496303470},
	number = {1},
	urldate = {2025-04-29},
	journal = {SIAM Journal on Optimization},
	author = {Lagarias, Jeffrey C. and Reeds, James A. and Wright, Margaret H. and Wright, Paul E.},
	month = jan,
	year = {1998},
	pages = {112--147},
}

@inproceedings{xu_optimo_2015,
	address = {New York, NY, USA},
	series = {{HRI} '15},
	title = {{OPTIMo}: {Online} {Probabilistic} {Trust} {Inference} {Model} for {Asymmetric} {Human}-{Robot} {Collaborations}},
	isbn = {978-1-4503-2883-8},
	shorttitle = {{OPTIMo}},
	url = {http://doi.acm.org/10.1145/2696454.2696492},
	doi = {10.1145/2696454.2696492},
	urldate = {2019-02-08},
	booktitle = {Proceedings of the {Tenth} {Annual} {ACM}/{IEEE} {International} {Conference} on {Human}-{Robot} {Interaction}},
	publisher = {ACM},
	author = {Xu, Anqi and Dudek, Gregory},
	year = {2015},
	pages = {221--228},
}

@article{abbasi_investigation_2024,
	title = {An investigation of perceived risk dimensions in acceptability of shared autonomous vehicles, a mediation-moderation analysis},
	volume = {14},
	copyright = {2024 The Author(s)},
	issn = {2045-2322},
	url = {https://www.nature.com/articles/s41598-024-74024-0},
	doi = {10.1038/s41598-024-74024-0},
	language = {en},
	number = {1},
	urldate = {2025-04-25},
	journal = {Scientific Reports},
	author = {Abbasi, Mohammadhossein and Mamdoohi, Amir Reza and Ciari, Francesco and Sierpiński, Grzegorz},
	month = oct,
	year = {2024},
	pages = {23276},
}

@phdthesis{akash_reimagining_2020,
	type = {thesis},
	title = {Reimagining {Human}-{Machine} {Interactions} through {Trust}-{Based} {Feedback}},
	url = {https://hammer.purdue.edu/articles/thesis/Reimagining_Human-Machine_Interactions_through_Trust-Based_Feedback/12493007/1},
	language = {en},
	urldate = {2025-04-25},
	school = {Purdue University Graduate School},
	author = {Akash, Kumar},
	month = jun,
	year = {2020},
	doi = {10.25394/PGS.12493007.v1},
}

@article{muir_trust_1996,
	title = {Trust in automation. {Part} {II}. {Experimental} studies of trust and human intervention in a process control simulation},
	volume = {39},
	issn = {0014-0139},
	doi = {10.1080/00140139608964474},
	number = {3},
	urldate = {2025-01-27},
	journal = {Ergonomics},
	author = {Muir, Bonnie M. and Moray, Neville},
	month = mar,
	year = {1996},
	pages = {429--460},
}

@article{williams_computational_2023,
	title = {A {Computational} {Model} of {Coupled} {Human} {Trust} and {Self}-confidence {Dynamics}},
	volume = {12},
	url = {https://dl.acm.org/doi/10.1145/3594715},
	doi = {10.1145/3594715},
	number = {3},
	urldate = {2025-04-17},
	journal = {J. Hum.-Robot Interact.},
	author = {Williams, Katherine J. and Yuh, Madeleine S. and Jain, Neera},
	month = jun,
	year = {2023},
	pages = {39:1--39:29},
}

@inproceedings{akash_toward_2020,
	address = {New York, NY, USA},
	series = {{ICMI} '20},
	title = {Toward {Adaptive} {Trust} {Calibration} for {Level} 2 {Driving} {Automation}},
	isbn = {978-1-4503-7581-8},
	doi = {10.1145/3382507.3418885},
	urldate = {2021-02-13},
	booktitle = {Proceedings of the 2020 {International} {Conference} on {Multimodal} {Interaction}},
	publisher = {Association for Computing Machinery},
	author = {Akash, Kumar and Jain, Neera and Misu, Teruhisa},
	month = oct,
	year = {2020},
	pages = {538--547},
}

@article{sato_automation_2020,
	title = {Automation trust increases under high-workload multitasking scenarios involving risk},
	volume = {22},
	issn = {1435-5566},
	doi = {10.1007/s10111-019-00580-5},
	language = {en},
	number = {2},
	urldate = {2024-11-20},
	journal = {Cognition, Technology \& Work},
	author = {Sato, Tetsuya and Yamani, Yusuke and Liechty, Molly and Chancey, Eric T.},
	month = may,
	year = {2020},
	pages = {399--407},
}

@article{stapel_-road_2022,
	title = {On-road trust and perceived risk in {Level} 2 automation},
	volume = {89},
	issn = {13698478},
	doi = {10.1016/j.trf.2022.07.008},
	language = {en},
	urldate = {2023-01-23},
	journal = {Transportation Research Part F: Traffic Psychology and Behaviour},
	author = {Stapel, Jork and Gentner, Alexandre and Happee, Riender},
	month = aug,
	year = {2022},
	pages = {355--370},
}

@article{hu_computational_2019,
	title = {Computational {Modeling} of the {Dynamics} of {Human} {Trust} {During} {Human}–{Machine} {Interactions}},
	volume = {49},
	issn = {2168-2305},
	doi = {10.1109/THMS.2018.2874188},
	number = {6},
	urldate = {2024-05-08},
	journal = {IEEE Transactions on Human-Machine Systems},
	author = {Hu, Wan-Lin and Akash, Kumar and Reid, Tahira and Jain, Neera},
	month = dec,
	year = {2019},
	pages = {485--497},
}

@article{melnicuk_effect_2021,
	title = {Effect of cognitive load on drivers’ {State} and task performance during automated driving: {Introducing} a novel method for determining stabilisation time following take-over of control},
	volume = {151},
	issn = {0001-4575},
	shorttitle = {Effect of cognitive load on drivers’ {State} and task performance during automated driving},
	doi = {10.1016/j.aap.2020.105967},
	urldate = {2024-11-12},
	journal = {Accident Analysis \& Prevention},
	author = {Melnicuk, Vadim and Thompson, Simon and Jennings, Paul and Birrell, Stewart},
	month = mar,
	year = {2021},
	pages = {105967},
}

@article{sonoda_displaying_2017,
	title = {Displaying {System} {Situation} {Awareness} {Increases} {Driver} {Trust} in {Automated} {Driving}},
	volume = {2},
	copyright = {https://ieeexplore.ieee.org/Xplorehelp/downloads/license-information/OAPA.html},
	issn = {2379-8904, 2379-8858},
	doi = {10.1109/TIV.2017.2749178},
	number = {3},
	urldate = {2024-11-14},
	journal = {IEEE Transactions on Intelligent Vehicles},
	author = {Sonoda, Kohei and Wada, Takahiro},
	month = sep,
	year = {2017},
	pages = {185--193},
}

@article{lee_trust_1994,
	title = {Trust, self-confidence, and operators' adaptation to automation},
	volume = {40},
	issn = {1071-5819},
	doi = {10.1006/ijhc.1994.1007},
	number = {1},
	urldate = {2023-12-18},
	journal = {International Journal of Human-Computer Studies},
	author = {Lee, John D. and Moray, Neville},
	month = jan,
	year = {1994},
	pages = {153--184},
}

@article{stapel_automated_2019,
	title = {Automated driving reduces perceived workload, but monitoring causes higher cognitive load than manual driving},
	volume = {60},
	issn = {13698478},
	doi = {10.1016/j.trf.2018.11.006},
	language = {en},
	urldate = {2024-11-13},
	journal = {Transportation Research Part F: Traffic Psychology and Behaviour},
	author = {Stapel, Jork and Mullakkal-Babu, Freddy Antony and Happee, Riender},
	month = jan,
	year = {2019},
	pages = {590--605},
}

@article{azevedo-sa_real-time_2021,
	title = {Real-{Time} {Estimation} of {Drivers}’ {Trust} in {Automated} {Driving} {Systems}},
	volume = {13},
	issn = {1875-4805},
	url = {https://doi.org/10.1007/s12369-020-00694-1},
	doi = {10.1007/s12369-020-00694-1},
	language = {en},
	number = {8},
	urldate = {2025-02-04},
	journal = {International Journal of Social Robotics},
	author = {Azevedo-Sa, Hebert and Jayaraman, Suresh Kumaar and Esterwood, Connor T. and Yang, X. Jessie and Robert, Lionel P. and Tilbury, Dawn M.},
	month = dec,
	year = {2021},
	pages = {1911--1927},
}

@article{hu_dynamic_2024,
	title = {Dynamic and quantitative trust modeling and real-time estimation in human-machine co-driving process},
	volume = {106},
	issn = {1369-8478},
	url = {https://www.sciencedirect.com/science/article/pii/S1369847824002006},
	doi = {10.1016/j.trf.2024.08.001},
	urldate = {2025-02-04},
	journal = {Transportation Research Part F: Traffic Psychology and Behaviour},
	author = {Hu, Chuan and Huang, Siwei and Zhou, Yu and Ge, Sicheng and Yi, Binlin and Zhang, Xi and Wu, Xiaodong},
	month = oct,
	year = {2024},
	pages = {306--327},
}

@article{perello-march_how_2024,
	title = {How {Do} {Drivers} {Perceive} {Risks} {During} {Automated} {Driving} {Scenarios}? {An} {fNIRS} {Neuroimaging} {Study}},
	volume = {66},
	issn = {0018-7208},
	shorttitle = {How {Do} {Drivers} {Perceive} {Risks} {During} {Automated} {Driving} {Scenarios}?},
	url = {https://doi.org/10.1177/00187208231185705},
	doi = {10.1177/00187208231185705},
	language = {en},
	number = {9},
	urldate = {2024-11-12},
	journal = {Human Factors},
	author = {Perello-March, Jaume and Burns, Christopher G. and Woodman, Roger and Birrell, Stewart and Elliott, Mark T.},
	month = sep,
	year = {2024},
	pages = {2244--2263},
}

@inproceedings{li_no_2019,
	address = {Utrecht Netherlands},
	title = {No {Risk} {No} {Trust}: {Investigating} {Perceived} {Risk} in {Highly} {Automated} {Driving}},
	isbn = {978-1-4503-6884-1},
	shorttitle = {No {Risk} {No} {Trust}},
	url = {https://dl.acm.org/doi/10.1145/3342197.3344525},
	doi = {10.1145/3342197.3344525},
	language = {en},
	urldate = {2023-01-24},
	booktitle = {Proceedings of the 11th {International} {Conference} on {Automotive} {User} {Interfaces} and {Interactive} {Vehicular} {Applications}},
	publisher = {ACM},
	author = {Li, Mengyao and Holthausen, Brittany E. and Stuck, Rachel E. and Walker, Bruce N.},
	month = sep,
	year = {2019},
	pages = {177--185},
}

@article{ji_gao_extending_2006,
	title = {Extending the decision field theory to model operators' reliance on automation in supervisory control situations},
	volume = {36},
	issn = {1558-2426},
	doi = {10.1109/TSMCA.2005.855783},
	number = {5},
	journal = {IEEE Transactions on Systems, Man, and Cybernetics - Part A: Systems and Humans},
	author = {Ji Gao and Lee, J.D.},
	month = sep,
	year = {2006},
	pages = {943--959},
}
